# A Technical Policy Blueprint for Trustworthy Decentralized AI

Hasan Kassem[1*], Orion Banks[2], Omar Benjelloun[3], Sergen Cansiz[4], Brandon Edwards[5], Patrick Foley[6], Inken Hagestedt[7], Taeho Jung[8], Peter Kairouz[3], Marco Lorenzi[4], Peter Mattson[3], Prakash Moorthy[1], Ann K Novakowski[2], Michael O'Connor[9], Bruno Rodrigues[10], Holger Roth[9], Micah Sheller[6], Dimitris Stripelis[6], Renato Umeton[1], Marc Vesin[4], Wenbin Zhang[11], Mic Bowman[5*], Alexandros Karargyris[1*]

[1]MLCommons, [2]Sage Bionetworks, [3]Google, [4]Inria, [5]Intel, [6]Flower, [7]Apheris
[8]University of Notre Dame, [9]NVIDIA, [10]Factored, [11]Florida International University

Contributions: Contributors in alphabetical order by last name except for contributors with * who initiated this effort and led this paper

Abstract: Decentralized AI can play a critical role in further unlocking AI asset marketplaces (e.g., healthcare data marketplaces). However this necessitates a robust, scalable governance layer. This paper tackles the challenge posed by inconsistent, bespoke policies that currently limit trust and interoperability among stakeholders. We are proposing a Technical Policy Blueprint centered on community-driven, machine-readable 'policy-as-code' objects. Our blueprint separates the complex logic of policy verification from the act of enforcement. The Policy Engine verifies required evidence (e.g., signatures, identification, payment proofs, etc). and issues a capability package. Asset Guardians then simply verify and apply this package, without needing to be reconfigured when policies change. This core concept of decoupling policy processing from capabilities is the key to creating a system that is transparent, auditable, and resilient to change, paving the way for trustworthy decentralized AI.

# 1. Introduction

AI asset marketplaces (e.g., data, models, workflows, compute) are becoming more ubiquitous. This trend is supported by industry (e.g., Datavant and Lifebit) and public initiatives (e.g., ARPA-H INDEX, European Health Data Space and NHS Federated Data Platform). These marketplaces facilitate AI asset transactions between stakeholders to perform a plethora of tasks such as AI/ML training, analytics, inference, benchmarking and evaluation creating new value (e.g., models, insights and intelligence).

AI asset marketplaces are part of a broader effort to develop infrastructure for decentralized AI, in which AI applications are developed and deployed using assets from multiple owners whose trust in one another may not already be established. In such an environment, improved scalable solutions for governance are critical to increasing participation by helping to communicate the needs of participants and establish trust that those will be met.

## 1.1 Problem Definition

Unlocking the full potential of AI asset governance requires a robust technical foundation—one that ensures reliable enforcement of policies around access control, monitoring, and asset integrity. Such a foundation is essential for fostering transparency and trust in AI asset marketplaces.

A significant barrier to achieving this is the prevalence of bespoke "policy languages" within organizations. These internal frameworks are often developed to answer critical questions such as: Who within the company can access my data? What are they permitted to do with it? How does their usage benefit me or my business unit? While these questions may summarize all concerns at the time of framework design, a more general community-driven approach to policy definition and enforcement can improve interoperability and scalability. Standardization can help to enable governance for all nuances of AI asset usage; addressing the who, what, where, when, how, and why of access and application. Importantly, data and AI governance are bundled together when we discuss AI governance below.

The gaps in governance frameworks due to a lack of standardization are further exacerbated by the complexity of establishing trust across diverse participants and systems. Verifying identities, credentials, and claims at scale remains a formidable challenge, especially amid inconsistent standards and fragmented compliance regimes.

Moreover, the development, management, and enforcement of policies at scale is both technically demanding and resource-intensive. Leveraging verifiable, tamper-proof mechanisms for policy enforcement can reduce dependence on institutional or personal trust. When compliance is demonstrable through transparent, cryptographically verifiable means, the need for human intermediaries diminishes—paving the way for scalable, trustworthy governance with integrity.

## 1.2 Existing AI Governance Tools in Decentralized AI Frameworks

### NVIDIA FLARE Site Policies

In NVIDIA FLARE, governance is decentralized through site policies[1] [2], which allow each participating site to independently enforce its own rules for resource usage, security, and privacy. These policies—defined through configuration files—govern how local resources are managed, who has access to what actions, and how sensitive information is protected. At runtime, the FLARE client automatically applies these configurations, ensuring that all operations comply with the site's local governance boundaries. This federated authorization framework enables organizations to collaborate more securely while maintaining control over their own compliance and data privacy constraints, and for example, implement job review mechanisms[3].

---

[1] NVIDIA. (2025). *Site policy management*. NVIDIA FLARE 2.7.0 Documentation. Retrieved from https://nvflare.readthedocs.io/en/main/user_guide/admin_guide/security/site_policy_management.html

[2] Kersten, K., & Dogra, P. (2022, October 25). *Federated learning from simulation to production with NVIDIA FLARE*. NVIDIA Technical Blog. Retrieved from https://developer.nvidia.com/blog/federated-learning-from-simulation-to-production-with-nvidia-flare/

## Flower Governance

The Flower Framework [4] [5] is designed to maximize ease of use, while putting appropriate checks in place for asset owners to decide what code gets executed on their infrastructure. Flower SuperGrid[6], the enterprise platform built on Flower, additionally brings decentralized policy enforcement defined at the site-level. Each organization granularly decides on the span of applications they trust, specific federations to take part in, and which datasets can be accessed. In highly sensitive enterprise environments, Flower SuperNodes[7] ensure that applications have been vetted and signed by a trusted entity, categorically limiting the potential for untrusted workloads to be launched in sensitive environments. Furthermore, audit logging provides best-in-class tracking of executions and communications down to the individual gradient. The combination of these capabilities grants each organization governance over their assets and a robust platform to demonstrate execution integrity for regulatory compliance.

## Fed-BioMed Governance

Fed-BioMed is designed to foster trustworthiness and tight collaboration among federation nodes, while enabling fast-turnaround medical research in a privacy-preserving manner[8]. It provides governance capabilities by allowing federated nodes to add, review, configure or revoke datasets at any time for federated training[9]. Federated learning (FL) workflows to be run on these datasets are required to be approved by the federated node before execution[10]. Fed-BioMed provides tools for inspection, approval, revocation and monitoring, so data owners can control and oversee FL execution.

## Apheris Governance

Computational Governance enables Data Custodians to enforce security, confidentiality, and auditability at the computation level. Asset Policies are rules set by the Data Custodian that define the allowed operations by named users on specified datasets.

---

[3] Boernert, E., Chmiel, J., & Antczak, L. (2023, September 28). Preventing health data leaks with federated learning using NVIDIA FLARE. NVIDIA Technical Blog. Retrieved from https://developer.nvidia.com/blog/preventing-health-data-leaks-with-federated-learning-using-nvidia-flare/

[4] Flower Labs. (2025). Flower Framework. [Source code]. GitHub. https://github.com/adap/flower

[5] Beutel DJ, Topal T, Mathur A, Qiu X, Fernandez-Marques J, Gao Y, Sani L, Li KH, Parcollet T, De Gusmão PP, Lane ND. Flower: A friendly federated learning research framework. arXiv preprint arXiv:2007.14390. 2020 Jul 28.

[6] Flower Labs. (2025). Flower Enterprise. https://flower.ai/enterprise/

[7] Flower Labs. (2025). Flower Architecture. https://flower.ai/docs/framework/explanation-flower-architecture.html

[8] Francesco Cremonesi, Marc Vesin, Sergen Cansiz, Yannick Bouillard, Irene Balelli, et al.. Fed-BioMed: Open, Transparent and Trusted Federated Learning for Real-world Healthcare Applications. Federated Learning Systems, 832, Springer Nature Switzerland, pp.19-41, 2025, Studies in Computational Intelligence, ff10.1007/978-3-031-78841-3_2ff. ffhal-04081557

[9] Fed-BioMed. (2025) *Deploying Datasets in Nodes* Documentation v6.2.0. Retrieved from https://fedbiomed.org/latest/user-guide/nodes/deploying-datasets/

[10] Fed-BioMed. (2025) *Training Plan Security Manager* Documentation v6.2.0. Retrieved from https://fedbiomed.org/latest/user-guide/nodes/training-plan-security-manager/

More specifically, Asset Policies define:

- who can perform computations on which dataset;
- which computations are allowed;
- parameter boundaries for these computations;
- the type of privacy controls to add at runtime (optional).

Compute Specifications (Compute Spec) are defined by the data scientist, who specifies what computation they intend to run on which data. This includes the particular model and parameters for securely running statistics functions and machine learning models on specified datasets.

The Apheris Gateway validates, automatically provisions, and executes Compute Specs within the Data Custodians environment. The aggregation of results happens securely on the Orchestrator. Sensitive data never leaves the Data Custodian's environment.

## 1.3 AI Governance in User Platforms

### Synapse Platform

Synapse, operated by Sage Bionetworks, aims to empower biomedical researchers with strong data governance through fine-grained permissions that reflect consent and policy-based conditions on data re-use (e.g., ethics board requirements or contractual terms). Additional safeguards are configured for controlled datasets that can be applied at multiple levels and inherited across asset hierarchies to ensure consistent enforcement. Metadata, provenance, and reporting features record the data lifecycle, across research assets, to support transparency and reproducibility. These same mechanisms can be extended to model sharing with gated access to certain artifacts (e.g., weights or parameters),.  enabling model execution while minimizing model risk.

## 1.4 AI Governance in Dataset Standards

### Croissant Standard

Croissant[11] is an open, community-built, standardized metadata vocabulary for ML datasets based on schema.org. Croissant describes key properties of ML datasets,  enabling data interoperability across ML frameworks and user platforms. Designed with extensibility in mind, the latest specification of Croissant enables governance capabilities by supporting  the Open Digital Rights Language (ODRL). Data creators and data owners can describe governance requirements in ODRL within their Croissant dataset description.

[11] Akhtar M, Benjelloun O, Conforti C, Foschini L, Giner-Miguelez J, Gijsbers P, Goswami S, Jain N, Karamousadakis M, Kuchnik M, Krishna S. Croissant: A metadata format for ml-ready datasets. Advances in Neural Information Processing Systems. 2024 Dec 16;37:82133-48.

# 2. Solution: A Technical Blueprint for Policies

Following the problem definition above, our collaborative effort in decentralized AI aims to:

1. Increase AI governance transparency through community-driven policy objects (i.e. machine-readable 'policy-as-code' templates that describe and encode AI governance requirements).
2. Enable tamper-proof policy enforcement.
3. Reduce infrastructure complexity by separating AI asset access/use policies processing from application-level policies.

Figure 1 shows this flow. Our approach aims at helping infrastructure providers keep development and maintenance overhead low while simultaneously offering transparency and ease-of-use of AI governance requirements described in machine-readable policy objects adopted by the community.

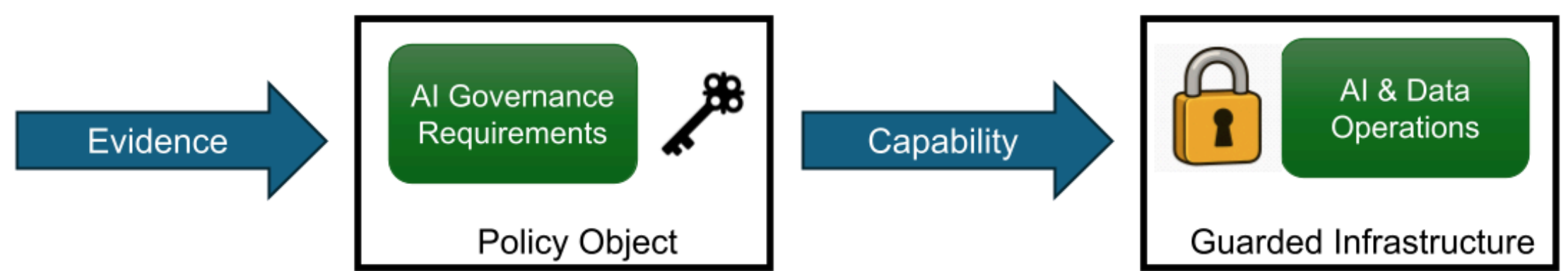

Figure 1. Computational policy enforcement unlocks capabilities on guarded AI assets

## 2.1 What does our Blueprint propose?

Returning to the trend of AI asset marketplaces and the key role of decentralized AI in enabling them, there is a need for a governance layer that can (1) interface with existing applications governance mechanisms, and (2) address policy gaps that existing applications cannot capture — for example, operational-level access controls such as granting data or model access conditional on payment or legal agreements.

Our collaborative technical blueprint proposes policy definitions, policy enforcement and policy limitations that can increase transparency, trust and community adoption, further unlocking decentralized AI.

Specifically our blueprint builds on these principles by:

- Defining a common policy abstraction, through policy objects, that expresses AI governance requirements in a consistent, machine-readable form.
- Establishing a collaborative process for identifying, codifying, and maintaining policy templates—allowing the community to evolve governance collectively
- Proposing a reference architecture for enforcing these policy objects across decentralized AI systems.

The blueprint aims to create that abstraction layer on top of existing governance efforts, promoting compliance with industry-specific rules and regulations, ensuring interoperability of policy code definitions, providing community-driven plug-and-play policy design templates, and enabling transparent, auditable, and tamper-evident tracking of policy usage and enforcement.

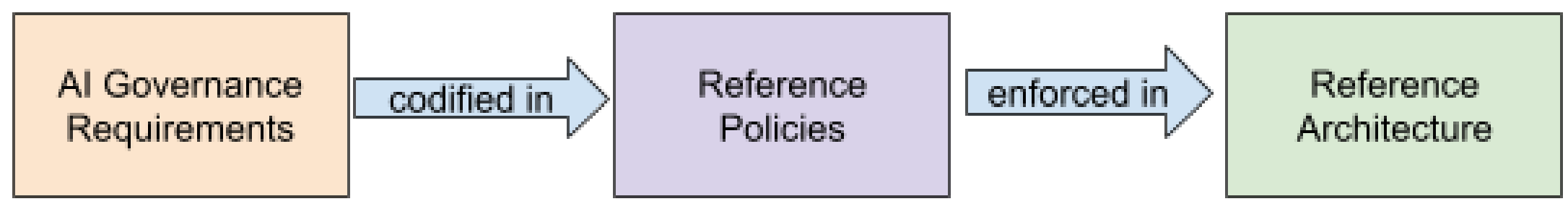

Figure 2. Our technical policy blueprint demonstrates how governance requirements are described in policies and enforced in the proposed reference architecture.

# 3. AI Governance Requirements

Following Figure 2, in an effort to better capture the specifications of our technical policy blueprint we surveyed the literature[12,13,14] to collect possible AI governance requirements that helped us map to broad reference policy requirements and reference architecture requirements. We managed to consolidate AI governance requirements into three (3) areas:

- Accountability & Oversight
  ↳Policies enabling operators of AI assets to assign roles and responsibilities
    ↳Technical infrastructure enforces accountability
- Protection & Integrity
  ↳Policies describing protection, security and privacy of AI assets
    ↳Technical infrastructure enforces integrity and confidentiality of AI assets during storage and runtime (note: policy enforcement engine should also be protected)
- Transparency & Monitoring
  ↳Policies describing provenance, monitoring and auditing capabilities on AI assets
    ↳Technical infrastructure enforces discoverability, traceability and persistence

The example below captures these areas:

A policy on a dataset may require signatures from designated data owners before allowing an operation (e.g., download). This represents an Accountability & Oversight requirement, ensuring accountability for the dataset's use.

This policy may be extended with a Protection & Integrity requirement, such as enforcing that the dataset can only be processed in a secure compute environment.

[12] National Institute of Standards and Technology. (2023). Artificial Intelligence Risk Management Framework (AI RMF 1.0) (NIST AI 100-1). U.S. Department of Commerce. Retrieved from https://nvlpubs.nist.gov/nistpubs/ai/nist.ai.100-1.pdf

[13] Organisation for Economic Co-operation and Development. (2019). Recommendation of the Council on Artificial Intelligence. Retrieved from https://legalinstruments.oecd.org/en/instruments/oecd-legal-0449

[14] European Commission. (2019). Ethics guidelines for trustworthy AI. High-Level Expert Group on Artificial Intelligence. Retrieved from https://digital-strategy.ec.europa.eu/en/library/ethics-guidelines-trustworthy-ai

In addition, a Transparency & Monitoring requirement might specify that every download and computation of AI assets be logged and discoverable, ensuring their provenance is captured, to enable auditing of dataset usage and track its derivatives for forensic traceability and accountability, such as licensing requirements.

# 4. Reference Policies

Policy-as-code are machine-readable policy objects that can be versioned, tested, and enforced computationally[15]. Typical policies cover security, compliance, governance and are evaluated by engines at different stages. In AI space policy-as-code can provide clear, testable rules that govern data, models, compute, training, and inference being executed in automated systems.

Codifying AI governance requirements into policy-as-code objects enables automation, consistency, unique interpretability, and minimized human error, ensuring that AI assets are used according to the intended AI governance requirement across diverse infrastructures and projects. Existing policy languages such as the Open Digital Rights Language (ODRL)[16], Rego (the policy language for Open Policy Agent[17]) or XACML (eXtensible Access Control Markup Language)[18] can be used to codify AI governance requirements into human-readable policy-objects.

Our blueprint (Figure 2) consists of having open community working streams to propose, develop, test and maintain policy-as-code objects based on focus areas (e.g., medical imaging AI and drug discovery AI). Through an open review process, we aim to provide transparency and enable the community to adopt and reuse these policy objects in their own projects or efforts. Specifically, our process follows open source lifecycle principles:

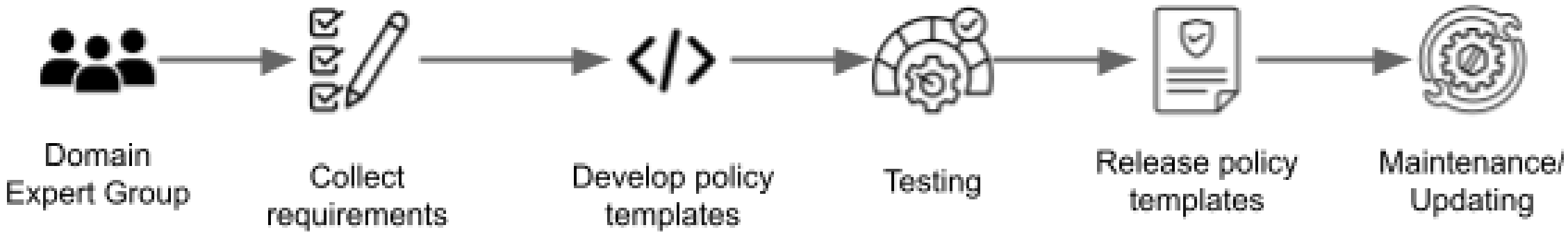

Figure 3. Collaborative Policy Development: Working Stream (domain experts) -> Collect AI Governance Requirements -> Codify requirements into policy-as-code templates -> Test and Review policy-object templates -> Release policy templates -> Maintain, Update and Archive

[15] Wei, H., Madhavji, N., & Steinbacher, J. (2025). Understanding everything as code: A taxonomy and conceptual model. arXiv. https://doi.org/10.48550/arXiv.2507.05100
[16] World Wide Web Consortium. (2018, February 15). *ODRL Information Model 2.2* [W3C Recommendation]. Retrieved from https://www.w3.org/TR/odrl-model/
[17] Open Policy Agent. (n.d.). *Policy language*. Retrieved December 6, 2025, from https://www.openpolicyagent.org/docs/policy-language
[18] Organization for the Advancement of Structured Information Standards (OASIS). (2013, January 22). *eXtensible Access Control Markup Language (XACML) Version 3.0* [OASIS Standard]. Retrieved from https://docs.oasis-open.org/xacml/3.0/xacml-3.0-core-spec-os-en.html

# 5. Reference Architecture

Our proposed reference architecture follows the AI governance flow definition (Figure 1) and aims to enforce policies via proof of evidence. Figure 4 highlights this flow at a high level. A user wants to perform an operation on a policy-protected AI asset (e.g., data access, model training or analytics):

1. First they must collect proof of evidence as described by the asset's policy object
2. Then, they provide this proof of evidence to the policy engine. The asset policy object managed by the policy engine will verify the evidence and give the user back a capability approval
3. Finally, the user provides this capability approval to the "Guardian environment" where the protected asset is hosted. The Guardian environment will only allow the operation on the protected asset if it is given a capability approval issued by its associated asset policy object.

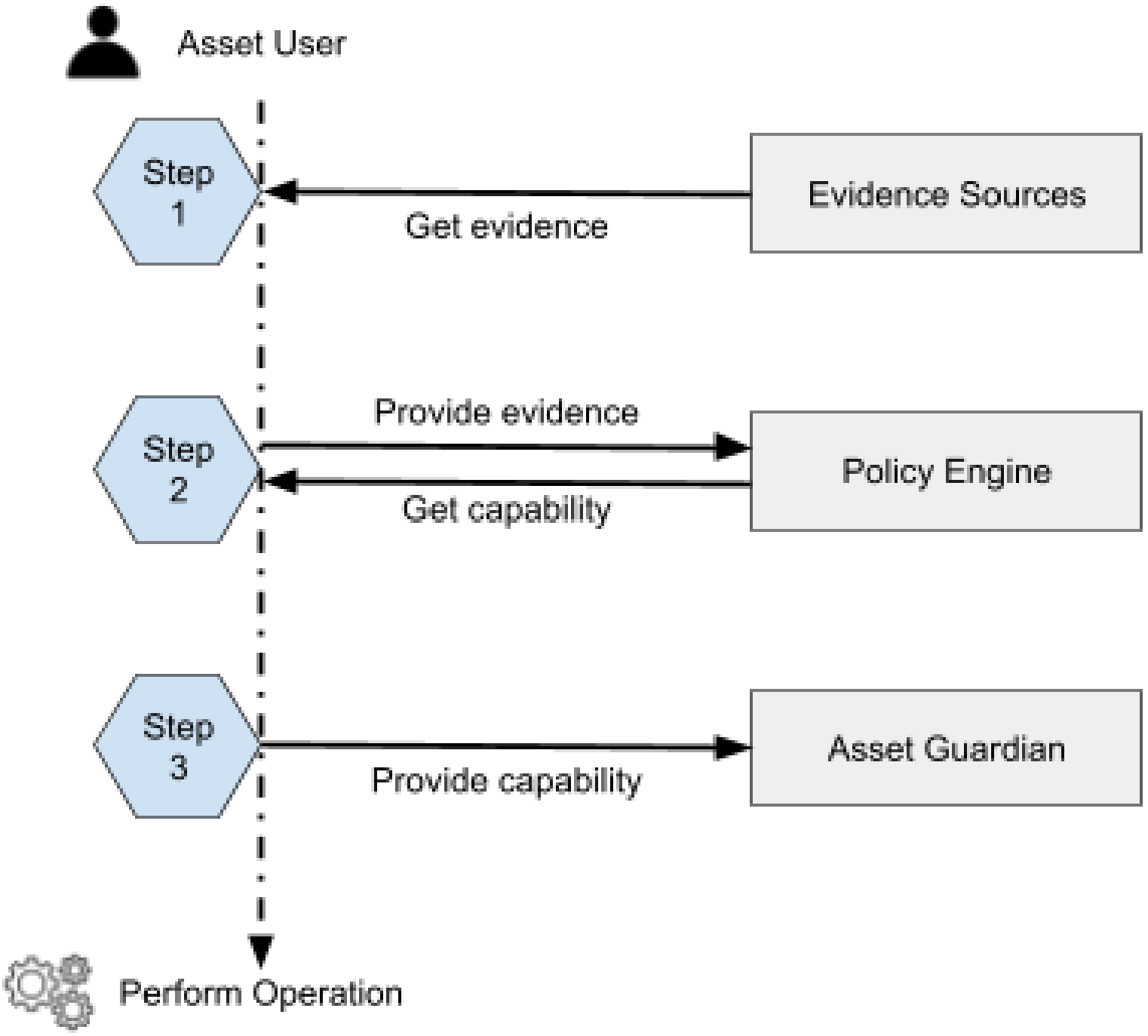

Figure 4. Reference Architecture with components enforcing AI Governance

To meet these requirements, we have identified the following components:

1. ***Evidence Sources***: Sources for evidence (e.g., trusted hardware evidence, payment proof or signatures by designated individuals), which asset users will provide for the *policy engine* to get a capability to access to a protected asset.

2. ***Policy Engine***: Provides the necessary backend to enforce policies by executing policy objects. The policy engine must be able to execute the policy objects with integrity and auditability, i.e., it should be able to keep an audit trail for interactions with policy objects and show that it executes policy objects as intended. The policy engine enforces policies by collecting required evidence from asset users, and upon evidence verification, a policy object issues a capability package to the asset users. A capability package represents the approval for performing a certain operation on the protected asset and will be consumed by the Guardian of the protected asset.

3. ***Guardians***:

a. ***Asset Guardians***: Environments that act as a gate for a protected asset. Asset guardians only act according to what a capability package issued by the corresponding policy object dictates.

b. ***Operation Guardians***: Environments where operations occur as intended (e.g., a TEE that ensures inference happens without malicious intervention or a machine managed by a certain trusted organization). These guardians are used when the operation itself needs to be controlled, not just the asset, such as in the case where the integrity of an inference workload is required. In this case, the inference will happen inside the operation guardian, and the operation guardian will communicate with the asset guardians to retrieve the involved assets and perform inference.

Additionally, users (i.e. asset owners, asset users, and reviewers) can interact with the aforementioned components through the following three (3) utility components:

1. ***Policy Manager*:** A client tool for asset owners that enables them to manage policy objects for their AI assets (e.g., create policy objects, register, update, revoke, delete, and archive)
2. ***Operations Manager***: A client tool for asset users that enables access to policy-protected assets; it facilitates the necessary interactions (Figure 4) with the policy engine and the guardians.
3. ***Auditing Manager***: A client tool for auditors that enables them to perform auditing and transparency reviews (e.g., verify actions and track access/usage of AI assets for compliance or for establishing accountability)

## 5.1 Decoupling “Policy Processing” from “Capabilities”

Within the policy engine, policy objects issue capability packages that define what actions are permitted. These capability packages are then consumed by asset guardians, which enforce access or execution permissions.

An *asset guardian* (e.g., Dataset guardian or Model guardian) does not need to understand the internal logic of a policy; it just requires trusting the policy object that issued the capability package. As long as the guardian is configured to recognize that trusted issuer, it simply verifies and applies the capabilities it receives.

This separation allows policy updates to occur independently: if an asset owner modifies a policy, the guardian does not need to be redeployed or reconfigured. This is the key contribution of the proposed reference architecture.

# 6. Policy-enabled AI Asset Examples

## 6.1 A Simple Example

### Context and Governance Requirements

A hospital ("*Dataset provider*") wants to contribute data to a federated clinical study. The *Dataset provider* requires a) signatures from designated individuals, and b) a download fee, before enabling download access to an asset user (e.g., a researcher).

### Asset Owner - Step 1: Dataset Policy Authoring

The *Dataset provider* will author the policy object ("*Dataset Policy*") such that it expects the designated individuals' signatures and a download fee as evidence. The *Dataset provider* will register the policy to the policy engine.

### Asset Owner - Step 2: Dataset Protection

Then, the dataset will be placed in an asset guardian ("*Data Guardian*"), which will only provide access to the dataset if a capability package issued by the *Dataset Policy* is provided.

### Asset User - Step 1: Evidence Preparation

When an asset user wants to download the dataset they will communicate with the designated individuals to get their signatures so they can download the dataset. The asset user will also pay the download fee (e.g., using a payment service the *Dataset provider* expects). Upon payment, the asset user will receive a payment receipt. In this way the asset user has acquired the two required pieces of evidence: signatures and payment receipt.

### Asset User - Step 2: Capability Package Acquisition

The asset user submits a request to the *Dataset Policy* object, which was registered on the policy engine by the *Dataset provider,* along with two required pieces of evidence. The *Dataset Policy* object will verify that:

- The signatures came from the designated individuals.
- The payment receipt is indeed issued by a trusted payment service, and it references the correct amount.

Then, the *Dataset Policy* object extracts the asset user identity from both the signatures and the payment receipt, verifies that the identities match, and issues a capability package to the asset user bearing the ability for this user to download the dataset from the *Data Guardian*. The user's identity is embedded in the capability package.

## Asset User - Step 3: Dataset Download

The asset user submits a request to the *Data Guardian* along with the capability package. The *Data Guardian* receives the capability package, verifies that it has been issued by the corresponding *Dataset Policy* object, and finally, allows the asset user to download the dataset. The *Data Guardian* transmits the asset to the asset user through a secure channel established by the identity of the asset user embedded in the capability package received.

## Extending Simple Example with Protection & Integrity Requirements

The previous example can be further extended to include the scenario where the *Dataset provider* requires, in addition to the signatures and the payment fee, that the data must be used only within a trusted execution environment (TEE) hardware (i.e., Protection & Integrity requirements).

The workflow is similar, albeit with the following changes:

- The proof of evidence is composed of three (3) pieces:
    - The signatures of the designated owners, as before
    - The payment receipt, as before
    - An attestation token issued by a certain trusted attestation service that asserts that the receiving hardware is a TEE.
- The *Dataset Policy* object now verifies the three (3) evidence pieces, and embeds the TEE hardware identity from the attestation token into the capability package, instead of using the asset user identity. This ensures that the data can only be downloaded within the TEE.
- The *Data Guardian*'s functionality doesn't change; it will still use the identity embedded in the capability package to establish a secure channel when sending the dataset, but in this scenario the identity will be that of the TEE, not the user.

This example also highlights the key contribution of the reference architecture: evidence -> capability package -> operation. In this way the policy processing is separated from the enforcement point in the data guardian. Therefore, when the *Dataset provider* updates their policy, the data guardian does not have to be re-deployed; only the policy object needs to be updated through the policy manager client tool.

Note: The purpose of this example is to demonstrate the overall workflow. It does not include more complex requirements or details. For example, as part of the TEE evidence verification by the policy object, the policy logic should have restrictions on the TEE attestation token, such as requiring that the attestation token contains signed claims that the TEE a) cannot be accessed by secure shell (SSH), b) doesn't have open ingress ports, and c) runs a certain verifiable piece of code on the dataset.
These mentioned restrictions can also be enforced similar to the example, but we omitted such details to keep the example simple and reader-friendly.

## 6.2 A Federated Learning Example

### Context and Governance Requirements

A hospital ("Dataset provider") wants to allow researchers to learn from their patients' clinical data; however, for compliance with local legislation, the data cannot leave the hospital's IT environment. Therefore, the data provider allows computations using a set of federated learning algorithms that they have evaluated as being sufficiently secure and privacy-preserving. For example, they require federated averaging and disallow federated stochastic gradient descent (SGD). In other words, the clients must train their local models sufficiently before sending their updated model weights to the server. Federated averaging is less vulnerable to privacy attacks against the training data during the learning process. In this use case, the data provider also requires at least one more data provider to join the federation.

### Asset Owner - Step 1: Dataset Policy Authoring

With the help of the policy manager, the asset owner authors the policy object. They consider domain knowledge about their data to select the most common and useful federated learning algorithms for their data type.
The data provider also registers their dataset in a computation environment where asset users can execute the FL client code upon request.

### Asset User - Step 1: Evidence Preparation

The asset user browses the registered datasets' metadata and identifies data of interest. They check the requirements for access and utilize the operations manager to gather the necessary evidence. For example, they upload their research proposal in PDF format. Moreover, the asset user selects the FL algorithm from the set of predefined algorithms based on their research goal and data.
The asset user repeats the process for at least one more dataset, as they want to use federated learning. Notice that the requirements of another asset owner may differ; the operations manager guides the asset user on how to submit the correct evidence.

### Asset User - Step 2: Capability Package Acquisition

A trusted third party verifies the asset user's research intent and ensures compliance with the asset owner's requirements for use.
The verification of the algorithmic requirements, such as the correct choice of algorithm and hyperparameters, as well as the presence of sufficiently many clients, is done automatically.
The operations manager coordinates all capability packages and makes them available for the data guardian.

### Asset User - Step 3: Federated Learning Execution

Once the asset user has gathered all capability packages, they can start the federated research project.

They submit one or several computation requests, providing the same capability package with each request. They inspect the output of each federated computation to inform their next steps; for example, failure of the FL algorithm to converge will result in different parameter or algorithmic choices.

### Extension of Requirements

The above example can be extended with additional requirements for security and privacy.
The asset owner could allow customizations of the federated learning algorithm if they are deemed secure by a third-party auditor or static code analysis checks. That allows the asset user more flexibility without compromising data security.
The asset owner could also prescribe a simulation of privacy attacks, and if the learned model is vulnerable to these attacks, it cannot be used. For example, if the learned model is susceptible to membership inference attacks, the asset owner's training data is at risk, and the asset user needs to change the learning regime. Differential privacy is the gold standard for protecting against membership inference attacks and may be used by the asset user or directly required by the asset owner.

# 7. Policy Engine

Policy objects are code. The policy engine is responsible for running this code, with the following required properties:

- Prove the integrity of policies processing (i.e., running policy objects)
- Trace user interactions with policy objects
- Ensure scalability with no single-point-of-failure
- Provide a standard interface: standard evidence input, standard capability package output, and standard policy object definitions.

A single centralized policy engine can enforce integrity and traceability by executing policy objects within a trusted environment and logging all executions. In this design, correctness depends entirely on the trustworthiness of the hosting entity. However, this assumption becomes problematic in open environments such as a universal AI asset marketplace. A single server also creates scalability bottlenecks and introduces a single point of failure.

To address scalability and eliminate central-trust assumptions, policy execution logs can be maintained on a distributed ledger. Distributed ledgers replicate state across multiple nodes and validate updates using consensus. This ensures immutability, transparency, and fault tolerance—making them a strong fit for multi-party policy governance without reliance on a central authority.

While distributed ledgers protect integrity and traceability, they do not inherently protect the confidentiality of the policy object’s state, inputs, or outputs. To solve this, policy execution can be performed inside Trusted Execution Environments (TEEs). The ledger nodes verify attestation reports from the TEEs before committing execution results. This provides end-to-end confidentiality and verifiable execution integrity.

Designing the policy engine as a distributed ledger with TEE-based policy objects execution gives the following advantages:

- Eliminates reliance on a single centralized server for audits or policy enforcement.
- Removes single points of failure, improving system resilience.
- Preserves the integrity and confidentiality of the policy objects' state, inputs, and outputs.
- Enables verifiable policy execution, since participants can independently confirm that a policy was executed correctly without accessing sensitive data.
- Supports automated and traceable governance, as policy issuance and evidence verification events are immutably recorded on the ledger.

The Private Data Objects[19] is an open-sourced project that can function as a policy engine. Other proprietary policy engines include Collibra AI Governance software[20] and Policy Engine - Securiti[21].

[19] Bowman, M., Miele, A., Steiner, M., & Vavala, B. (2018). Private data objects: An overview (arXiv:1807.05686). arXiv. https://arxiv.org/pdf/1807.05686
[20] Collibra. (n.d.). *AI Governance*. Retrieved December 6, 2025, from https://www.collibra.com/products/ai-governance
[21] Securiti. (n.d.). *Policy Engine*. Retrieved December 6, 2025, from https://securiti.ai/products/policy-engine/